\documentclass[english,pra,showpacs,reprint,superscriptaddress]{revtex4-1}
\usepackage[latin9]{inputenc}
\usepackage{color}
\usepackage{babel, array, multirow}
\usepackage{verbatim}
\usepackage{bm}
\usepackage{amsmath}
\usepackage{amssymb}
\usepackage{graphicx}
\usepackage[unicode=true,pdfusetitle,bookmarks=true,bookmarksnumbered=false,bookmarksopen=false,breaklinks=false,pdfborder={0 0 0},pdfborderstyle={},backref=false,colorlinks=true]{hyperref}
\hypersetup{citecolor=blue}
\makeatletter
\usepackage{epstopdf}
\usepackage{amsfonts}
\usepackage{mathrsfs}\usepackage{bm}\usepackage{appendix}\usepackage{color}
\usepackage{txfonts}
\usepackage{float}
\usepackage{epstopdf}\usepackage{ulem}
\makeatother

\begin{document}

\title{Topologically protected Fano resonance in photonic valley Hall insulators}
\author{Chang-Yin Ji}
\affiliation{Key Lab of advanced optoelectronic quantum architecture and measurement (MOE), Beijing Key Lab of Nanophotonics $\&$ Ultrafine Optoelectronic Systems, and School of Physics, Beijing Institute of Technology, Beijing 100081, China.}
\affiliation{China Academy of Engineering physics, Mianyang, Sichuan, 621900, China.}
\author{Yongyou Zhang}
\email[Corresponding author: ]{yyzhang@bit.edu.cn}
\affiliation{Key Lab of advanced optoelectronic quantum architecture and measurement (MOE), Beijing Key Lab of Nanophotonics $\&$ Ultrafine Optoelectronic Systems, and School of Physics, Beijing Institute of Technology, Beijing 100081, China.}
\author{Bingsuo Zou}
\affiliation{Guangxi Key Laboratory of Processing for Non-ferrous Metals and Featured Materials, School of Physical science and Technology, Guangxi University, Nanning 530004, China.} 
\author{Yugui Yao}
\email[]{ygyao@bit.edu.cn}
\affiliation{Key Lab of advanced optoelectronic quantum architecture and measurement (MOE), Beijing Key Lab of Nanophotonics $\&$ Ultrafine Optoelectronic Systems, and School of Physics, Beijing Institute of Technology, Beijing 100081, China.}

\date{\today}
\begin{abstract}
Rapidly developing photonics brings many interesting resonant optical phenomena, in which the Fano resonance (FR) always intrigues researchers because of its applications in optical switching and sensing. However, its sensitive dependence on environmental conditions makes it hard to implement in experiments. We in this work suggest a topologically-protected FR based on the photonic valley Hall insulators, immune to the system impurities. The topologically-protected FR is achieved by coupling the valley-dependent topological edge states (TESs) with one double-degenerate cavity. The $\delta$-type photonic transport theory we build reveals that this topological FR dates from the interference of the two transmissions that are attributed to the parity-odd cavity mode and the parity-even one. We confirm that the induced Fano line shape of the transmission spectra is robust against the bending domain walls and disorders. Our work may provoke exciting frontiers for manipulating the valley transport and pave a way for the topologically protected photonic devices such as optical switches, low-threshold lasers, and ultra-sensitive sensors.
\end{abstract}

\maketitle

\section{Introduction}

Since Haldane and Raghu transferred electronic topological phases to classical photonics \cite{haldane2008possible}, photonic topological insulators have triggered plenty of research interests \cite{lu2014topological, sun2017two, wu2017applications, khanikaev2017two, xie2018photonics, ozawa2019topological, rider2019perspective,ota2020active}. They not only enrich the realm of photonics but also provide a promising research perspective for robustly manipulating photon transport against system impurities. Because the macroscopic optical meta-atoms can be easily engineered into various architectures, photonic metamaterials provide an ideal platform to explore and verify the fascinating topological phases that exist in the electronic systems \cite{klitzing1980new, haldane1988model, kane2005quantum, bernevig2006quantum, moore2010birth, qi2011topological, fu2011topological}. A number of optical analogue topological phases have been realized theoretically and experimentally, such as quantum Hall effect (QHE) \cite{wang2008reflection,yu2008one,wang2009observation,ao2009one,skirlo2014multimode,skirlo2015experimental,jacobs2015photonic,xiao2017photonic}, quantum spin Hall effect (QSHE) \cite{hafezi2011robust,khanikaev2013photonic,chen2014experimental,ma2015guiding,wu2015scheme,xu2016accidental,
cheng2016robust,he2016photonic,anderson2017unidirectional,yves2017crystalline,zhu2018topological,barik2018topological,yang2018visualization,
gorlach2018far,peng2019probing,PhysRevResearch}, Weyl semimetals \cite{lu2013weyl,lu2015experimental,bravo2015weyl,lin2016photonic,gao2016photonic,chen2016photonic,chang2017multiple,
wang2017optical,yang2018ideal,wang2019photonic,cerjan2019experimental,jia2019observation,
yang2019spontaneous,ye2019photonic,yin2019tunable}, high-order topological insulators \cite{peterson2018quantized,li2018topological,xie2018second,mittal2019photonic,el2019corner,xie2019visualization,
chen2019direct,ota2019photonic,zhang2019higher}, photonic topological valley Hall effect \cite{deng2014observation,ma2016all,gao2017valley,chen2017valley,dong2017valley,wu2017direct,kang2018pseudo,yang2018topological,
chen2018tunable,wu2018reconfigurable,ni2018spin,noh2018observation,chen2018valley,gao2018topologically,xue2018spin,
song2019valley1,proctor2019manipulating,orazbayev2019quantitative,
ma2019topological,wang2019tunable,xu2020topological,yamaguchi2019gaas,shalaev2019robust, he2019silicon,chen2019valley, deng2019vortex, chen2019all, wu2019interlayer,Yamaguchi_2019,Yoshimi20}, and so on \cite{rechtsman2013photonic, gao2015topological, lu2016symmetry, slobozhanyuk2017three, mittal2019photonic1, yang2019realization}. Topological quantum states can also be induced by spatially uniform optical excitations \cite{lubatsch2019evolution, lubatsch2019behavior}.

One remarkable application of the photonic topological insulators
is the robust transport of the topological edge states (TESs), protected by the nontrivial topology of the systems. The robust transport has been widely studied in experiments and theories, while the effect of defect-based cavities on the robustness of TESs is still far less considered \cite{ji2019transport, ji2020fragile, ota2020active}. For the optical analogues of Chern topological insulators, the chiral TESs are indeed unprecedented robustness. However, the TES transport may show fragility in the systems with the time-reversal (TR) symmetry (TRS) due to a pair of time-reversal modes can always be coupled together by a possible mechanism that breaks the unidirectional transport of the TESs, for example, recent works on the QSHE in the photonic and acoustic topological insulators \cite{ji2019transport, ji2020fragile}. The breaking of the unidirectional transport provides the tunability for the transport of the TESs, possessing significant potential applications, such as switches. Unfortunately, the topological insulators in Ref.~\cite{ji2019transport} not only require $C_{6}$ symmetry, but also are hard for designing a large matching gap between two topologically-inequivalent insulators. Additionally, there is an inherent band gap at the TR invariant point in the TES dispersion, which indicates that the backscattering of the TES propagation is not totally immune around such a point. These characteristics are not favorable for potential applications. It is thus significant and urgent to find new topological systems to overcome them. 

The photonic valley Hall insulators (PVHIs) which can be achieved only by common dielectric materials provide a simple route that enable explorations of the photonic topological state in a wide range of electromagnetic spectrum
from microwaves to the visible. Here, we show that PVHIs can serve as an ideal candidate to implement the manipulation of the TESs, simultaneously without the above shortcomings. We find that the defect cavities with frequency falling into the bulk band gap of the PVHIs can break the backscattering immunity of the valley TES propagation protected by non-zero valley Chern number, leading to the perfect reflection \cite{ji2019transport}. Moreover, an ideal Fano resonance (FR), i.e., valley transmission sharply changes from zero to one, is implemented by a degenerate cavity that is felicitously engineered to side couple with the valley TESs. More importantly, the profile of this FR is robust against the bending domain wall and disorders. The tunability of the valley TES transport, together its topological protection against fabrication imperfection enlarge the realm of applications of the photonic topological insulators, because of high desire in many scenarios for designing topology-based devices  \cite{nozaki2010sub, wu2012fano, volz2012ultrafast, chua2011low, zhen2013enabling, khanikaev2013fano, stern2014fano, heeg2015interferometric, le2015enhanced, li2013ultrasensitive, zhao2015ultrafast, limonov2017fano, wang2017exciton}.

The rest of the present work is organized as follows. In Sec.~\ref{PVHIs}, the PVHIs are designed and their topologies are confirmed by the band Berry curvatures. In Sec.~\ref{Fano}, the Fano line-shape transmission is first demonstrated by coupling the valley TESs with a degenerate cavity and then revealed to be protected by the system topology. In Sec.~\ref{theory}, a topological waveguide-cavity transport theory is build to demonstrate the underlying mechanism of the fascinating FR. At last, we arrive at a conclusion in Sec.~\ref{conclusion}.

\begin{figure}
\centering{}\includegraphics[width=0.48\textwidth]{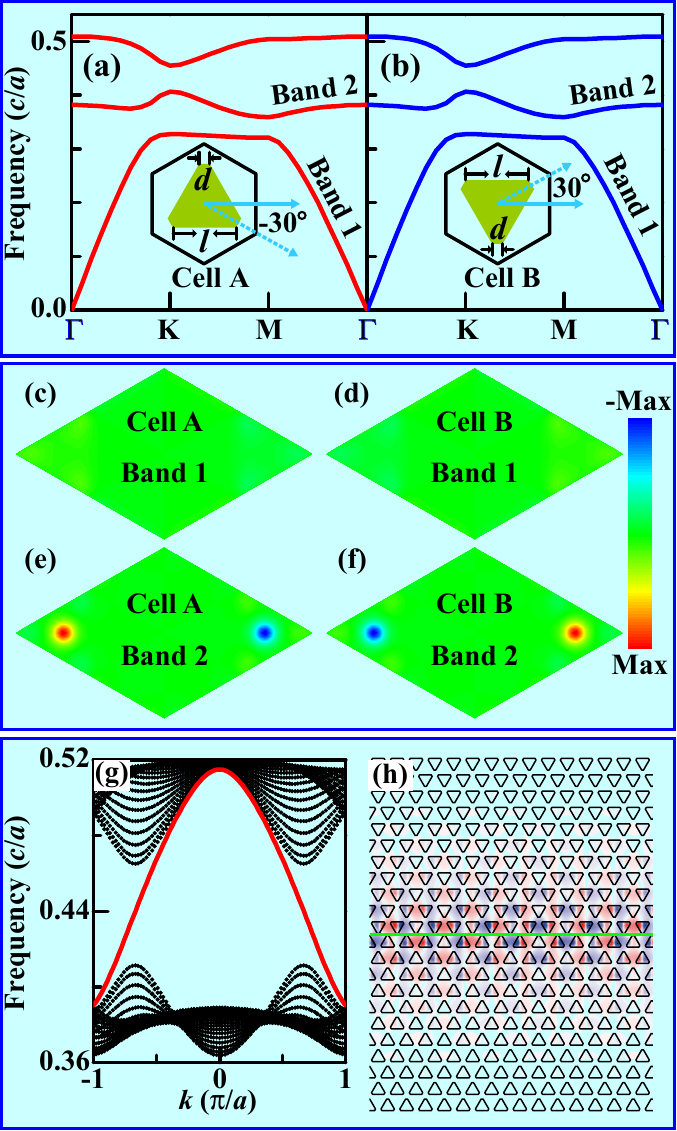}
\caption{(a, b) Band structures of two photonic topological valley Hall insulators. Their unit cells A and B are, respectively, given in the insets of (a) and (b), both comprised of an angle-cut triangle meta-atom embedded in air. Their lattice periods are set to be $a$ with $l=0.61a$ and $d=0.1a$. The triangular-like meta-atom is nonmagnetic material with relative dielectric constant $\varepsilon_r=16$. (c-f) Distributions of lattice field strength in the first Brillouin zone. (c) and (d) are those of bands 1 for cells A and B, while (e) and (f) are those of bands 2 for cells A and B. (g) One dimensional photonic band diagram of the ribbon-shaped super cell in (h). The red and black curves are for topological edge states and bulk states, respectively. (h) Magnetic field distribution of the topological edge states. The attenuated field tells that the edge state is localized around the domain wall.}
\label{structures}
\end{figure}

\section{Photonic valley Hall insulators}\label{PVHIs}

The designed PVHIs are the two-dimensional hexagonal lattices with an angle-cut triangle meta-atom embedded in the air background, see the unit cells A and B in Figs.~\ref{structures}(a) and \ref{structures}(b). The long and short side lengths are set to be $l=0.61a$ and $d=0.1a$, respectively, where $a$ is the lattice constant. For them the meta-atoms are composed of nonmagnetic material with relative dielectric constant $\varepsilon_r=16$ which can be achieved in microwave and light \cite{kang2018, PRL121024301}. The magnetic and electric fields for the considered photonic topological states are perpendicular and parallel to the unit-cell plane, respectively. Since the triangle meta-atoms are rotated $\pm30^\circ$ with respect to the horizontal lattice vector direction for the cells A and B in the Figs.~\ref{structures}(a) and \ref{structures}(b), respectively, their six Dirac degeneracy points at the Brillouin zone (BZ) corners are both lifted up to form a global band gap between the second and third bands, essentially owing to the breaking of the system $C_{3v}$ symmetry. The mutual mirror symmetry along the horizontal lattice vector between the unit cells A and B is responsible for the identity of the two bands in Figs.~\ref{structures}(a) and \ref{structures}(b), but note that the two bands present different topology near the Brillouin zone (BZ) corners, referred to Figs.~\ref{structures}(c)-\ref{structures}(f).

The different topologies of the first two bands can be confirmed by analyzing the Berry curvature, defined as \cite{fukui2005chern}
\begin{equation}
F^{(n)}_{\bm k}\equiv\nabla_{\bm k}\times \bm A^{(n)}_{\bm k},
\end{equation}
with the band index $n$, Bloch wave vector $\bm k$, and Berry connection $\bm A^{(n)}_{\bm k}$. The Berry connection has the following form,
\begin{equation}
\bm A^{(n)}_{\bm k}=\left\langle u^{(n)}_{\bm k}\right|\nabla_{\bm k}\left|u^{(n)}_{\bm k}\right\rangle,
\label{BC}
\end{equation}
where $\left|u^{(n)}_{\bm k}\right\rangle$ is the periodic part of the Bloch wave function in the system. It is more convenient to calculate the lattice field strength of $\widetilde{F}^{(n)}_{\bm k}$ rather than $F^{(n)}_{\bm k}$ \cite{fukui2005chern}. As the discretized area of $d^2\bm k$ is small enough, one has $\widetilde{F}^{(n)}_{\bm k}\approx F^{(n)}_{\bm k}d^2\bm k$. The values of $\widetilde F^{(n)}_{\bm k}$ for the bands 1 in Figs.~\ref{structures}(a) and \ref{structures}(b) are shown in Figs.~\ref{structures}(c) and \ref{structures}(d), respectively. The zero $\widetilde F^{(n)}_{\bm k}$ in the whole BZ tells that the bands 1 are topologically trivial. On the contrary, the bands 2 in Figs.~\ref{structures}(a) and \ref{structures}(b) have non-zero $\widetilde F^{(n)}_{\bm k}$ near the BZ corners, see Figs.~\ref{structures}(e) and \ref{structures}(f), respectively. The non-zero $\widetilde F^{(n)}_{\bm k}$ implies the non-trivial topology of the bands 2. The TRS is responsible for the zero sum of $\widetilde{F}^{(n)}_{\bm k}$ in the whole BZ. This non-trivial topology can be characterized by the valley-dependent topological index, that is, the valley Chern number defined as $C_V\equiv C_{\bm K}- C_{\bm K'}$. The $C_{\bm K}$ ($ C_{\bm K'}$) is the sum of $\widetilde F^{(n)}_{\bm k}$ around the $\bm K$ ($\bm K'$)-valley. Figures \ref{structures}(e) and \ref{structures}(f) show that the sign of $C_{\bm K}$ ($C_{\bm K'}$) for the band 2 in Fig.~\ref{structures}(a) is opposite with that for the band 2 in Fig.~\ref{structures}(b). Therefore, there is a non-zero $C_V$ across the domain wall when it is formed by the PTVH insulators in Figs.~\ref{structures}(a) and \ref{structures}(b).

\begin{figure*}
\centering{}
\includegraphics[width=0.95\textwidth]{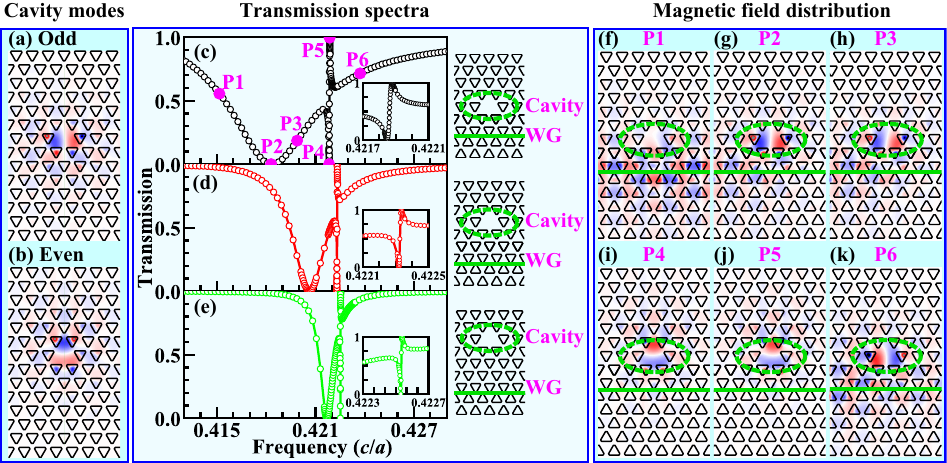}
\caption{(a, b) Field distributions of the two degenerate cavity modes with odd and even parities (along the horizontal direction), respectively. (c-e) Transmission spectra of the valley TESs coupled with the double-degenerate cavity calculated by the finite element method within the COMSOL code. The optical structures are adopted on the right side, whose distance between the cavity and the WG increases from 2 layers to 4 layers. The insets are the enlarged views of the Fano profiles and the solid lines are guides for eyes. (f-k) Field distributions whose frequencies of the incident light correspond to the magenta dots marked in (c). The positions of the domain walls and cavities are denoted by the green solid lines and dashed ellipses, respectively.}
\label{cavity}
\end{figure*}

Note that $|C_V|$ is in the range of $(0,\ 1)$, in general not a well-defined integer \cite{he2019silicon}. This is attributed to the large band gap that overlaps the Berry curvatures at the $\bm K$ and $\bm K'$ valleys. In fact, the existence of the TESs does not require a well-defined integer valley Chern number, see the red curves in Fig.~\ref{structures}(g) which demonstrates the dispersion of the TESs with the band gap in the range of (0.413, 0.463)$\frac{c}{a}$. The TES dispersions in Fig.~\ref{structures}(g) correspond to the domain wall in  Fig.~\ref{structures}(h),  which can be taken as an ideal waveguide (WG), denoted by the green solid line. The color map there displays an example of the field distribution of the TES. Note that the valley TESs can also be achieved in a square lattice \cite{zhu2019negative}. This abundance of the architecture diversity benefits the application of the valley topological insulators. Moreover, another merit of the present system is that the rightward- and leftward-moving TESs are fully gapless, see Fig.~\ref{structures}(g), with respect to the optical analogue of the QSHE in Ref.~\cite{ji2019transport}. If the designed defect cavities can tune the transportation of the valley TESs, the PVHIs would show widespread potential for the optical application, for example, the topologically-protected FR.

\section{Topologically-protected Fano resonance}\label{Fano}

For the topological phases in electronic systems, the propagation of the TESs is robust against the system imperfections, such as bending domain walls, defect cavities, disorders and so on. As a rational conjecture, researchers instinctively believe that the propagation of all types of the TESs in photonics should be immune to the aforementioned imperfections. However, the defect cavities with frequencies in the topological band gap can significantly break the robust pseudospin-polarized transport of photonic TESs, because the introduced defect cavities inevitably violate the system crystal symmetries \cite{ji2019transport}. This property as a merit provides a powerful tool for tuning the propagation of the TESs in various photonic topological devices. Since the valley topology depends on the crystal symmetry, it can also be achieved in the valley topological systems. In Figs.~\ref{cavity}(a) and \ref{cavity}(b), a double degenerate cavity is designed as an example to show the propagation tuning of the valley TESs. The cavity is formed by simply removing one photonic meta-atom. Since the cavity has the $D_3$ symmetry, it holds a pair of degenerate modes with the odd and even parities, respectively, see the field distributions in Figs.~\ref{cavity}(a) and \ref{cavity}(b). The eigenvalue analysis confirms that their frequencies are equal to ${\sim}0.42269c/a$, being in the topological band gap.

Let's focus on the strong interaction of such a type of degenerate cavities with the valley TESs in the following. When the cavity is side placed by the topological WG, it will strongly influence the propagation of the valley TESs, see the transmission spectra in Figs.~\ref{cavity}(c)-\ref{cavity}(e) whose structures are shown on the right side. The transmission spectra are all obtained by the scattering matrix method \cite{lu2017observation,ji2019transport}. The $D_3$ symmetry of the cavity is broken by the presence of the WG and therefore, the degeneracy of the cavity modes is lifted up, reflected by the two zero transmission points in frequency. The two zero transmission points confirm that the propagation of the valley-dependent TESs is strongly disturbed near the cavity eigenfrequencies. The interesting is that the fundamental Fano line shape is produced in the present topological architectures. The enlarged views of Fano profiles in the insets of Figs.~\ref{cavity}(c)-\ref{cavity}(e) tell that the FR is pretty ideal, since the transmission is ultra-sharp from zero to unity. The transmission spectra from Figs.~\ref{cavity}(c) to \ref{cavity}(e) also show that the zero transmission frequency point and the line width can be conveniently tuned through changing the coupling distance between the cavity and the topological WG.

The origin of the FR is uncovered by the field distributions in Figs.~\ref{cavity}(f)-\ref{cavity}(k) whose frequencies of the incident light beams correspond to the points of P1 to P6 in Fig.~\ref{cavity}(c), respectively. The wide transmission dip is attributed to the strong coupling between the odd cavity mode and the WG, see Figs.~\ref{cavity}(f)-\ref{cavity}(h) and \ref{cavity}(k) for P1-P3 and P6, while the Fano line shape is from the coupling of the even cavity mode with the WG, see Figs.~\ref{cavity}(i) and \ref{cavity}(j) for P4 and P5. The valley pseudospin scattering is mainly affected by the odd cavity mode, while the even cavity mode only plays a dominant role in a narrow frequency range around the Fano line shape. Thus, the coupling of the TES with the odd cavity mode is stronger than that with the even cavity one, which is attributed to the different mode matching of the odd and even cavity modes with the TES along the WG direction, see Figs.~\ref{cavity}(f)-\ref{cavity}(k). Consequently, the odd mode serves as a bright one, while the even mode serves as a dark one \cite{zhang2008plasmon-induced, lukyanchuk2010the, bochkova2018high-q}. The affections of the odd and even cavity modes concur to bring about the destructive and constructive interferences that create the Fano line shape transmission spectra, which will be confirmed by the $\delta$-type coupling theory we build in Sec.~\ref{theory}.

\begin{figure}
\centering{}\includegraphics[width=0.48\textwidth]{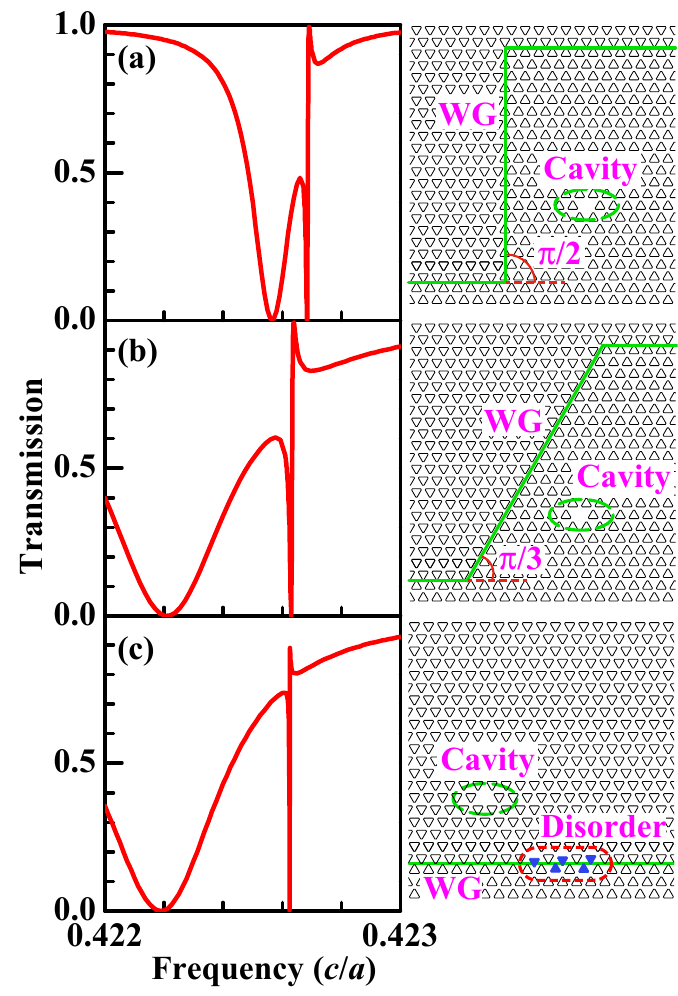}
\caption{(a, b) Transmission spectra of the valley-dependent edge states
coupled with side cavity with different bending angles for the domain walls. (c) Transmission spectrum of the coupling system under disorder. The corresponding structures are shown on the right side. The green lines indicate the domain walls and the green dashed ellipses indicate the position of the cavity. The red dashed oval denotes the disordered position.}
\label{robust}
\end{figure}

The FR is always interesting in the realm of photonics \cite{RevModPhys.82.2257, limonov2017fano, PhysRevBcore, PhysRevResearchto, liu2020fano}, and has broad applicabilities in prominent optical devices, such as optical switches, low threshold lasers, high quality factor filters, and ultra-sensitive sensors \cite{wu2012fano, chua2011low, zhen2013enabling, stern2014fano}. The inherent characteristic of the ultra-sharp asymmetric line shape is susceptible to the changes of the surrounding environment and geometry. Especially, such a sensitivity often makes it hard to achieve in the experiments. Overcoming these unfavorable factors commonly requires advanced manufacturing techniques, associated with large costs. The concept of the topological FR has been introduced into one dimensional periodic classical system where the unprecedented robustness to disorders is protected by the non-trivial Zak phase \cite{zangeneh2019topological}. Compared with the FR in the one dimensional system, the two-dimensional optical architecture with easy fabrication in Fig.~\ref{cavity} can show more potentials in optics. Moreover, the FR presented in Fig.~\ref{cavity} is also topologically protected but by the non-zero valley Chern number, different from the Zak phase in Ref.~\cite{zangeneh2019topological}. This topological protection is confirmed by the robustness of the FR to the bending of the WGs and disorders that are intentionally introduced into the system, see Fig.~\ref{robust}. The two WGs with the bending angles of $\pi/2$ and $\pi/3$ are shown in Figs.~\ref{robust}(a) and \ref{robust}(b), respectively, both which unequivocally tell that the Fano line shape is robust to the bending WGs. Note that the FR is also immune to other bending angles and multi-corner WGs \cite{lu2017observation}, though they are not shown here.

As one of the most common imperfections, the disorder is introduced by randomly perturbing the positions of some optical meta-atoms, see the red dashed oval in Fig.~\ref{robust}(c). Obviously, the corresponding spectrum demonstrates the existence of the FR, which is protected by the non-trivial valley topology. Another feature is the perfect reflection, i.e., the zero transmission at the P2 or P4 point, referred to Fig.~\ref{cavity}(c). Since it does not disappear under the bending of the WGs and disorder, it is also protected by the valley topology, see Fig.~\ref{robust}. Note that the perfect reflection is hard to be realized in those trivial waveguide-cavity systems in the presence of impurities. In contrast to previous studies on topologically protected high transmission, the transport tuning of the TESs by a side-coupled cavity is a hallmark feature for the crystal-symmetry-protected topological insulators, which is reflected by the perfect reflection and FR proposed in the PVHIs. They are both immune to the bending of the WGs and system disorders. The topologically protected FR, not only against the impurities but also with ideal characteristics, have many merits to serve as high-performance optical devices.

\begin{figure}
\centering{}\includegraphics[width=0.48\textwidth]{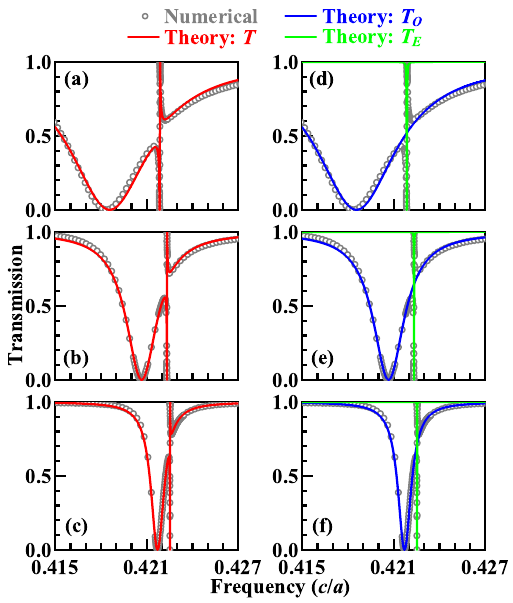}
\caption{(a-c) Theoretical fitting of the transmissivity $T$ (red solid curves) with respect to the numerical data (gray circular dots). (d-f) Transmissivities $T_O$ (blue curves) and $T_E$ (green curves) corresponding to the transmissivity $T$ in (a-c), respectively. The numerical data in (a-c) and (d-f) are identical with those in Figs.~\ref{cavity}(c-e), respectively.}
\label{fitting}
\end{figure}

\section{$\delta$-type coupling theory}\label{theory}

To uncover the coupling mechanism between the degenerate cavity and topological WG, a photonic transport theory is builded based on the quantum field method in this section. We start from the following field Hamiltonian,
\begin{align}
{\cal H} & = {\cal H}_{W} + {\cal H}_C + {\cal H}_I,
\label{eq:H}
\end{align}
where ${\cal H}_{W}$, ${\cal H}_C$, and ${\cal H}_I$ are the Hamiltonians for the WG, cavity, and their coupling. The WG Hamiltonian ${\cal H}_W$ reads \cite{shen2009theory,Shen20191,chengCoherent,Jia20131,JiangPhysRevA},
\begin{align}
{\cal H}_{W} & =\int dx\left[\hat{L}^{\dag}(x)\hat{\omega}\left(i\partial_x\right)\hat{L}(x)+{\hat{R}}^{\dag}(x)\hat{\omega}\left(-i\partial_x\right)\hat{R}(x)\right],\label{eq:HW}
\end{align}
where ${\hat R}^\dag(x)$ and ${\hat L}^\dag(x)$ $\left({\hat R}(x)\right.$ and $\left.{\hat L}(x)\right)$ express the creation (annihilation) operators of rightward- and leftward-moving TESs at the coordinate of $x$, respectively. The TES dispersion is linearized with the form of $\hat{\omega}\left(\pm i\partial_x\right)=\omega_0\pm iv_g \frac {\partial}{\partial x}$ with the corresponding intercept $\omega_0$ and group velocity $v_g$. The cavity Hamiltonian ${\cal H}_C$ reads,
\begin{align}
{\cal H}_C & =\omega_O\hat{c}_O^{\dag}\hat{c}_O+\omega_E\hat{c}_E^{\dag}\hat{c}_E,\label{eq:HC}
\end{align}
where $\hat{c}^\dag_O$ and $\hat{c}^\dag_E$ ($\hat{c}_O$ and $\hat{c}_E$) are the creation (annihilation) operators of the odd and even cavity modes with frequencies of $\omega_O$ and $\omega_E$, respectively. Note that there is a small splitting between $\omega_O$ and $\omega_E$ due to the breaking of the cavity symmetry by the domain wall. According to the previous work \cite{ji2019transport}, the coupling functions of the odd and even cavity modes with the TESs are also odd and even, respectively. Without loss of physics, we use the following Hamiltonian to describe the coupling between the two cavity modes and the valley TESs, i.e.,
\begin{align}
{\cal H}_I & = \int dx V_O\left[\delta(x+w)-\delta(x-w)\right]\left[\hat{R}^{\dag}(x)+\hat{L}^{\dag}(x)\right]\hat{c}_O\nonumber\\
& +\int dxV_E\delta(x)\left[\hat{R}^{\dag}(x)+\hat{L}^{\dag}(x)\hat{c}_E\right]+{\rm h.c.},\label{eq:HI}
\end{align}
where $V_O$ ($V_E$) is the coupling strength of the TESs with the odd (even) cavity mode. The width of $w$ measures the extension of the odd coupling function along the WG direction, while the extension of the even coupling function is neglected. The cavity position is taken as the original point of coordinate $x$. The complex coupling functions, of course, can work also, such as the Gaussian function (even) and its first-order derivative (odd), but they outcome a too complex result to understand the underlying physics \cite{ji2019transport}. This is just the reason why the $\delta$-type coupling functions are adopted in Eq.~\eqref{eq:HI}. Since they have caught the key information, i.e., the parity of the two cavity modes, they should be in agreement with the FR discussed above.

To find the system transmission, we use the following single-particle state,
\begin{align}
|\Phi\rangle & =\left({\cal C}_Oc_O^{\dag}+ {\cal C}_Ec_E^{\dag}\right)|\varnothing\rangle \nonumber\\
&+ \int dx\left[{\cal R}(x){\hat{R}}^{\dag}(x)+{\cal L}(x){\hat{L}}^{\dag}(x)\right]|\varnothing\rangle.
\label{eq:eig}
\end{align}
Here, ${\cal R}(x)$ and ${\cal L}(x)$ are the wave functions of the rightward- and leftward-moving TESs. ${\cal C}_O$ and ${\cal C}_E$ are the excitation amplitudes of the odd and even cavity modes. $|\varnothing\rangle$ represents the vacuum state with zero photons in the whole system. Substituting Eqs.~\eqref{eq:H} and \eqref{eq:eig} into the steady Schr\"odinger equation,
\begin{align}
 {\cal H}|\Phi\rangle=\omega|\Phi\rangle
  \label{eq:Sc}
\end{align}
we arrive at a set of coupled equations for ${\cal R}(x)$, ${\cal L}(x)$, ${\cal C}_O$, and ${\cal C}_E$, i.e.,
\begin{subequations}
\begin{align}
\omega{\cal R}(x)&=\hat{\omega}\left(-i\partial_x\right){R}(x)\nonumber\\ &+V_O\left[\delta(x+w)-\delta(x-w)\right]C_O+V_E\delta(x){C_E},\\
\omega{L}(x)&=\hat{\omega}\left(+i\partial_x\right){L}(x)\nonumber\\
&+V_O\left[\delta(x+w)-\delta(x-w)\right]C_O+V_E\delta(x){C_E},\\
\omega{C_O}&=\omega_O C_O+ V_O^\ast[R(-w)+L(-w)-R(w)-L(w)],\\
\omega{C_E}&=\omega_E C_E+ V_E^\ast\left[R(0)+L(0)\right],
\end{align}\label{eq:dyn}
\end{subequations}
with the frequency $\omega$ for the incident TES. Using the following wave functions ${\cal R}(x)$ and ${\cal L}(x)$,
\begin{subequations}
\begin{align}
{\cal R}(x)&=\theta(-x-w)e^{ikx}+t_1\theta(-x)\theta(x+w)e^{ikx}\nonumber\\
&\quad+t_2\theta(x)\theta(-x+w)e^{ikx}+t\theta(x-w)e^{ikx},\\
{\cal L}(x)&= r\theta(-x-w)e^{-ikx}+r_1\theta(-x)\theta(x+w)e^{-ikx}\nonumber\\
&\quad+r_2\theta(x)\theta(-x+w)e^{ikx},
\end{align}\label{eq:RL}
\end{subequations}
where $\theta(x)$ is the step function. $t$ and $r$ are the total transmission and reflection coefficients, respectively. Substituting Eq.~\eqref{eq:RL} into Eq.~\eqref{eq:dyn} gives $t$ as,
\begin{align}
t=1 &-i\frac{J_O\left[1-\cos(\phi)\right]}{\left[\omega-\omega_O+J_O\sin(\phi)\right]+iJ_O\left[1-\cos(\phi)\right]}\nonumber\\
&-i\frac{J_E}{\omega-\omega_E+iJ_E},
\label{eq:t}
\end{align}
where $J_O=v_g^{-1}|V_O|^2$, $J_E=v_g^{-1}|V_E|^2$, and $\phi=2kw$. The transmissivity reads $T=|t|^2$, which can well capture the FR of the system. The transmission coefficient can be divided into two types that account for the odd and even cavity modes. That is, the transmissivity $T_O=\left|t(J_E\equiv0)\right|^2$ $\left(T_O=\left|t(J_O\equiv0)\right|^2\right)$ provides the transmission spectrum of the TES just owing to the odd (even) cavity mode. As a result, the Fano line shape dates from the interference of these two types of the transmission.

\begin{table}
  \centering
  \caption{Fitting paramenters for Fig.~\ref{cavity}}\label{fittingparas}
  \renewcommand\arraystretch{1.5}
  \begin{tabular}{p{1.0cm}<{\centering}|p{2.2cm}<{\centering}|p{2.2cm}<{\centering}|p{2.2cm}<{\centering}}
  \hline\hline
   &  Fig.~\ref{cavity}(c)&  Fig.~\ref{cavity}(d) & Fig.~\ref{cavity}(e) \\
   \hline
   $\omega_O$ & $0.42146{c\over a}$ &  $0.42205{c\over a}$  & $0.42223{c\over a}$ \\
   \hline
   $\omega_E$ & $0.42186{c\over a}$ & $0.42233{c\over a}$  &  $0.42253{c\over a}$ \\
   \hline
   $J_O$ & $2.91378\times 10^{-3}{c\over a}$ &  $1.39751\times 10^{-3}{c\over a}$ & $5.23541\times 10^{-4}{c\over a}$ \\
   \hline
   $J_E$ & $1.47229 \times 10^{-5} {c\over a}$ & $7.02339\times 10^{-6}{c\over a}$  & $3.34527\times 10^{-6}{c\over a}$ \\
   \hline
   $\phi$ & $0.52795\pi$ & $0.46222\pi$  & $0.49946\pi$ \\
\hline\hline
\end{tabular}
\end{table}

This is confirmed by fitting the transmission spectra in Figs.~\ref{cavity}(c)-\ref{cavity}(e) with Eq.~\eqref{eq:t}, whose results are demonstrated in Figs.~\ref{fitting}(a)-\ref{fitting}(c), respectively, and the corresponding fitting parameters of $(\omega_O, \omega_E, J_O, J_E, \phi)$ are listed in the Table \ref{fittingparas}. Note that the gray circular dots in Figs.~\ref{fitting}(a)-\ref{fitting}(c) are those replicas of the numerical data shown in Figs.~\ref{cavity}(c)-\ref{cavity}(e), respectively. Since the red fitted curves are in agreement with the numerical calculations, the key physics of the FR is well captured by the established theory, though it has a simple form. The FR dates from the interference of the two kinds of the transmissions that are controlled by the parity-odd cavity mode and the parity-even one, respectively, which can be seen from Figs.~\ref{fitting}(d)-\ref{fitting}(f). The transmission spectra of $T_O$ have a wide transmission dip, coinciding with the global behavior of the numerical data, while those of $T_E$ have a far narrower transmission dip, coinciding with the FR positions. Their destructive and constructive interferences bring about the Fano line shape.

The fitting parameters in Table \ref{fittingparas} provide several important results. (i) The small frequency difference between the even and odd modes, i.e., $\omega_E - \omega_O$, roughly decreases with increasing the distance between the cavity and the WG, owing to the decrease of the breaking of the cavity symmetry $D_3$ by the domain wall. (ii) The coupling strengths of $J_O$ and $J_E$ also decrease with increasing the distance between the cavity and the WG. (iii) $J_O$ is about 200 times larger than $J_E$, reflecting the odd cavity mode as a bright one with respect to the even cavity mode as a dark one. (iv) The phase of $\phi\approx0.5\pi$ measures the extension of the odd coupling function along the WG direction, corresponding to $2w\approx0.65a$ which is consistent with the field distribution in Figs.~\ref{cavity}(g), \ref{cavity}(h), and \ref{cavity}(k). These qualitative analysis demonstrates that the topological FR originates from the interference of the two types of the transmission due to the coupling of the valley TESs with the odd and even cavity modes, described by Eq.~\eqref{eq:t}. Since the $\delta$-type coupling theory wonderfully catches the Fano line shape of the transmission spectra, it can provide a guidance for designing related topological optical devices. Moreover, such theory can also be used to describe the spatiotemporal evolution of the TESs, avoiding giant computing cost when simulating the TES transport in time domain with finite element method.

\section{conclusion}\label{conclusion}

In summary, the photonic topological valley Hall insulators as a special case of the TRS systems was used to reveal an inherent aspect that optical cavities with the resonance frequencies falling into the topological band gap
can completely destroy the robustness of the photonic TES transport. This phenomenon and the related applications are both far less investigated in the field of topological photonics. We demonstrated that an ideal Fano line shape for the transmission spectra can be obtained when the valley-dependent topological edge states couple with a double-degenerate cavity, behaving as the sharply changing of the transmission from zero to one. The Fano resonance is well in agreement with the $\delta$-type photonic transport theory we build. This theory reveals that the topological Fano resonance dates from the interference of the two transmissions that are controlled by the parity-odd cavity mode and the parity-even one. As a result, the Fano shape can be adjusted by controlling the coupling strengths of the TESs with the parity-odd and even cavity modes. In addition, we also demonstrated that the Fano resonance in this topological system is robust against the system impurities. Our work provides a realistic scheme to manipulate the valley transport and accordingly, paves the way for a broad applications of the photonic valley Hall insulators.

\section*{Acknowledgement}
We thank professor Jian-hua Jiang for the useful discussion on the work.
The authors also gratefully acknowledge financial support from the National Natural Science Foundation of China (Grant Nos. 11304015,  11734003) and
National Key R$\&$D Program of China (Grant No.  2016YFA0300600).

%\bibliographystyle{aipnum4-1}
%\bibliographystyle{elsarticle-num-names}
%\bibliography{ref}

\end{document}